\documentclass[11pt]{article}
\usepackage{graphicx}
\usepackage{hyperref}
\pdfoutput=1

\begin{document}



\title{Lagrangian Investigation of Auto-ignition in a Hydrogen Jet Flame in a Vitiated Co-flow:
Animations of Particle Trajectories in Composition Space from PDF Model Calculations}

\author{Haifeng Wang and Stephen B. Pope \\
\\\vspace{6pt} Sibley School of Mechanical and Aerospace
Engineering, \\ Cornell University, Ithaca, NY 14853, USA}

\maketitle

\begin{abstract}
PDF model calculations have been performed of the Cabra lifted
hydrogen flame in a vitiated co-flow.  Particle trajectories are
extracted from the Lagrangian particle method used to solve the
modeled PDF equation.  The particle trajectories in the mixture
fraction-temperature plane reveal (at successive downstream
locations): essentially inert mixing between the cold fuel jet and
the hot co-flow; the auto-ignition of very lean particles; and,
subsequent mixing and reaction, leading to near-equilibrium
compositions.  The purpose of this paper is to present animations of
the particle trajectories obtained using different turbulent mixing
models.
\end{abstract}

\section{Introduction}
\label{introduction}

We recently reported \cite{Wang-Pope-CTM-07} a PDF modeling study of
the H$_2$/N$_2$ lifted jet flame in a vitiated co-flow, which was
studied experimentally by Cabra {\em et al.}\cite{cabra02}. Particle
trajectories are extracted from the Lagrangian particle method used
to solve the modeled PDF equation. The particle trajectories in the
mixture fraction-temperature ($\xi$-$T$) plane reveal the dominant
processes occurring at different locations within the flame. The
purpose of this paper is to present some animations of the particle
trajectories obtained from these PDF calculations.

Full details of the PDF model and its numerical implementation are
given in \cite{Wang-Pope-CTM-07}.  Calculations are performed using
each of the three commonly-used turbulent mixing models.  These are:
the interaction by exchange with the mean (IEM) model
\cite{villermaux72} (or, equivalently, the least-mean-square
estimation (LMSE) model \cite{dopazo74}); the Euclidean minimum
spanning tree (EMST) model \cite{subramaniam98}; and, the modified
Curl (MC) model \cite{janicaka79,nooren97}.

\section{Results}


\begin{figure}
\centering
\includegraphics[width=0.75\textwidth]{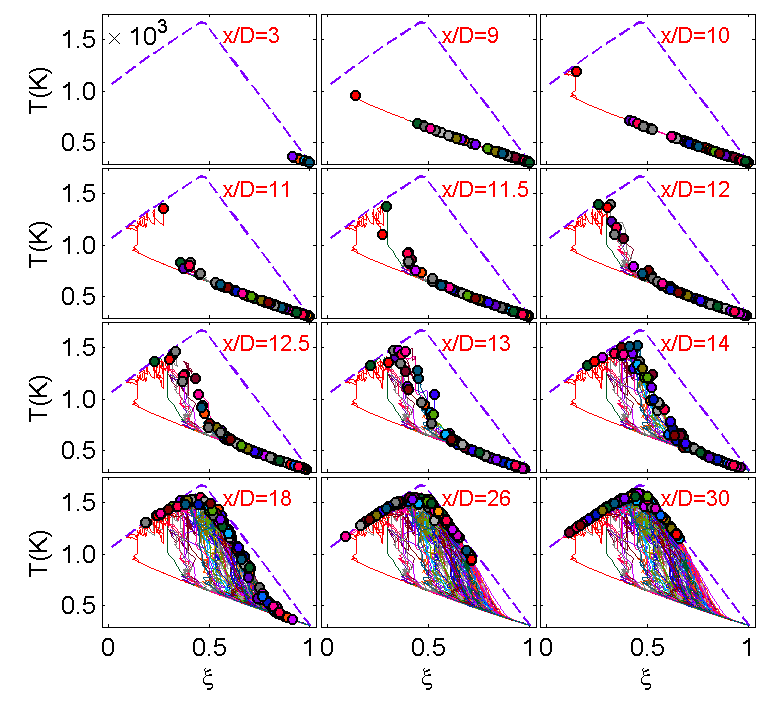}
\caption{Particle trajectories from the fuel region
in the Cabra lifted flame using the EMST model}
\label{TrajH2_Fuel_EMST}
\end{figure}

Figure \ref{TrajH2_Fuel_EMST} shows particle trajectories in the
$\xi$-$T$ plane at different axial locations, $x/D$.  In each plot:
pure fuel corresponds to the lower right corner,
$(\xi,T)\approx(1,300K)$; the co-flow stream corresponds to the
middle left, $(\xi,T)\approx(0,1033K)$; and the dashed line
corresponds to chemical equilibrium.  The trajectories shown are for
particles originating in the fuel stream.

Initially, ($x/D\leq 9$) the particles move in the plane exclusively
by mixing. A particle trajectory due to pure mixing is a nearly
straight line between the cold fuel temperature and the hot co-flow
temperature. Pure mixing yields a partially premixed mixture of fuel
and oxidizer at different mixture fractions. At about $x/D=10$, some
particles near the oxidizer side start to ignite first: the ignition
delay time is shortest for these very lean particles ($\xi \approx
0.05$).

After the rapid auto-ignition of the first few particles, these
relatively hot burnt particles (at $x/D>11$ in Fig.
\ref{TrajH2_Fuel_EMST}) mix with adjacent particles in composition
space, thus raising their temperature (and radical concentration)
and hence promoting their auto-ignition. Therefore the ignition
progressively moves to richer mixtures. This burning process is not
exclusively the auto-ignition of the particles. Both reaction and
mixing play important roles.  Eventually ($x/D\geq 26$) all
particles come close to chemical equilibrium.  Further discussion on
the processes involved is given in \cite{Wang-Pope-CTM-07}.

\begin{figure}[t]
\centering
\includegraphics[width=0.75\textwidth]{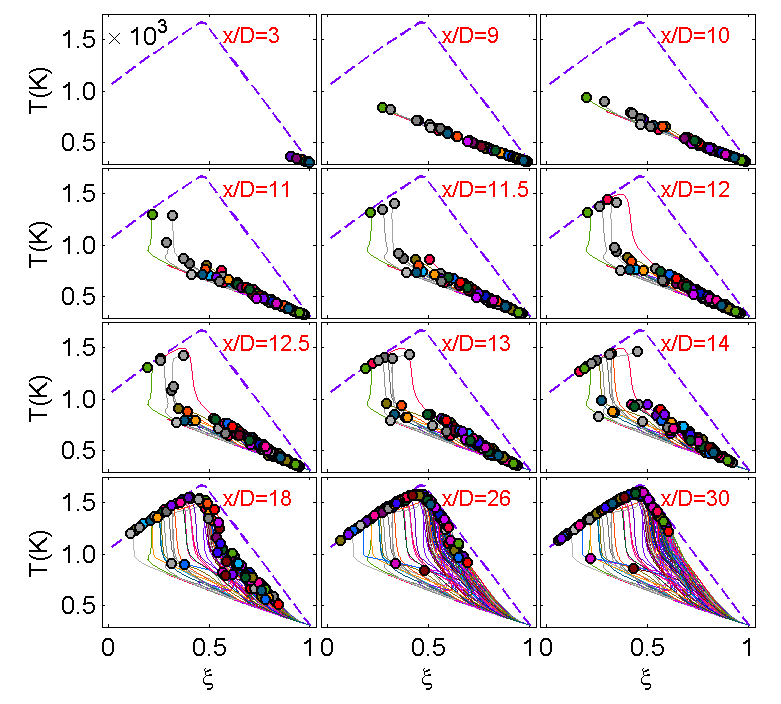}
\caption{Particle trajectories from the fuel region in the Cabra
lifted flame using the IEM model} \label{TrajH2_Fuel_IEM}
\end{figure}

\begin{figure}[t]
\centering
\includegraphics[width=0.75\textwidth]{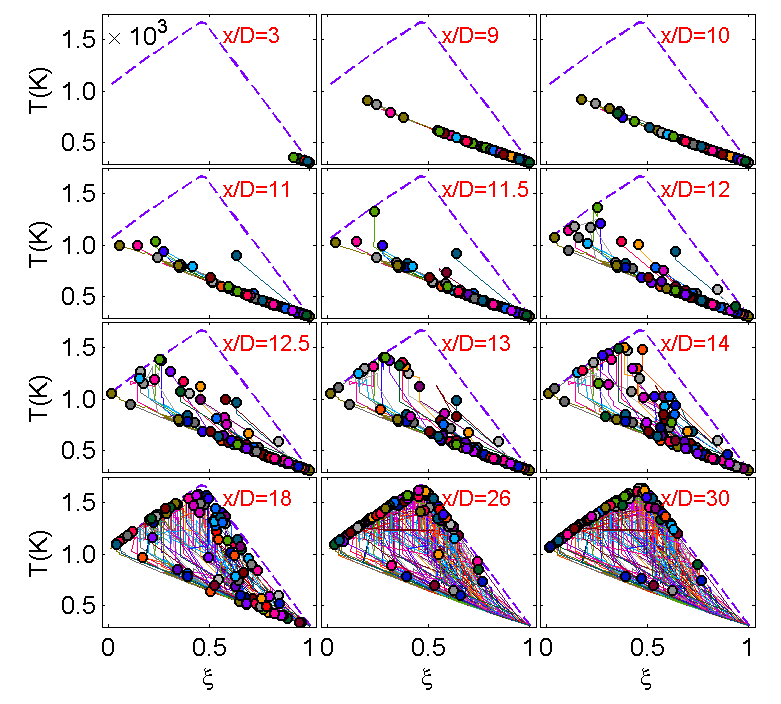}
\caption{Particle trajectories from the fuel region in the Cabra
lifted flame using the MC model} \label{TrajH2_Fuel_MCurl}
\end{figure}

Figures \ref{TrajH2_Fuel_IEM} and \ref{TrajH2_Fuel_MCurl} and show
the corresponding trajectories given by the IEM and MC mixing
models.

\href{http://ecommons.library.cornell.edu/bitstream/1813/8237/2/LIFTED_H2_EMST_FUEL.mpg}{Video
1} and
\href{http://ecommons.library.cornell.edu/bitstream/1813/8237/4/LIFTED_H2_IEM_FUEL.mpg}{video
2} are animations, showing the evolution of the particles in the
$\xi$-$T$ plane as they move downstream, according to the EMST and
IEM models, respectively. In the animations, all of the tracked
particles are shown (about 1,200), whereas the figures show just 100
trajectories selected randomly.


\section*{Acknowledgements}
This work is supported by the Air Force Office of Scientific
Research, Grant FA9550-06-1-0048 and by Department of Energy, Grant
DE-FG02-90ER.  This research was conducted using the resources of
the Cornell Theory Center, which receives funding from Cornell
University, New York State, federal agencies, foundations, and
corporate partners.

\end{document}